\documentclass[aps,prl,reprint,superscriptaddress,amsmath,amssymb,amsfonts,showpacs]{revtex4-1}
\usepackage{graphicx}% Include figure files
\usepackage{dcolumn}% Align table columns on decimal point
\usepackage{bm}% bold math
\usepackage{hyperref}% add hypertext capabilities
\usepackage{breakurl}
\usepackage{latexsym}
\usepackage{color}

\begin{document}

% Use the \preprint command to place your local institutional report
% number in the upper righthand corner of the title page in preprint mode.
% Multiple \preprint commands are allowed.
% Use the 'preprintnumbers' class option to override journal defaults
% to display numbers if necessary
%\preprint{}

%Title of paper
\title{Direct measurement of the mass difference of $^{163}$Ho and $^{163}$Dy solves $Q$-value puzzle for the neutrino mass determination}

% repeat the \author .. \affiliation  etc. as needed
% \email, \thanks, \homepage, \altaffiliation all apply to the current
% author. Explanatory text should go in the []'s, actual e-mail
% address or url should go in the {}'s for \email and \homepage.
% Please use the appropriate macro foreach each type of information

% \affiliation command applies to all authors since the last
% \affiliation command. The \affiliation command should follow the
% other information
% \affiliation can be followed by \email, \homepage, \thanks as well.
%\author{}
%\email[]{Your e-mail address}
%\homepage[]{Your web page}
%\thanks{}
%\altaffiliation{}
%\affiliation{}

\author{S. Eliseev}
\affiliation{Max-Planck-Institut f\"ur Kernphysik, Saupfercheckweg 1, 69117 Heidelberg, Germany}

\author{K. Blaum}
\affiliation{Max-Planck-Institut f\"ur Kernphysik, Saupfercheckweg 1, 69117 Heidelberg, Germany}

\author{M. Block}
\affiliation{GSI Helmholtzzentrum f\"ur Schwerionenforschung GmbH, Planckstra{\ss}e 1, 64291 Darmstadt, Germany}
\affiliation{Helmholtz-Institut Mainz, 55099 Mainz, Germany}
\affiliation{Institut f\"ur Kernchemie, Johannes Gutenberg-Universit\"at, 55128 Mainz, Germany}

\author{S. Chenmarev}
\affiliation{Max-Planck-Institut f\"ur Kernphysik, Saupfercheckweg 1, 69117 Heidelberg, Germany}
\affiliation{Physics Faculty of St.Petersburg State University, 198904, Peterhof, Russia}

\author{H. Dorrer}
\affiliation{Institut f\"ur Kernchemie, Johannes Gutenberg-Universit\"at, 55128 Mainz, Germany}
\affiliation{Paul Scherrer Institute, 5232 Villigen, Switzerland}
\affiliation{Universit\"at Bern, 3012 Bern, Switzerland}

\author{Ch.E. D\"ullmann}
\affiliation{GSI Helmholtzzentrum f\"ur Schwerionenforschung GmbH, Planckstra{\ss}e 1, 64291 Darmstadt, Germany}
\affiliation{Helmholtz-Institut Mainz, 55099 Mainz, Germany}
\affiliation{Institut f\"ur Kernchemie, Johannes Gutenberg-Universit\"at, 55128 Mainz, Germany}
\affiliation{PRISMA Cluster of Excellence, Johannes Gutenberg-Universit\"at, 55099 Mainz, Germany}

\author{C. Enss}
\affiliation{Kirchhoff Institut f\"ur Physik, Heidelberg Universit\"at, INF 227, 69120 Heidelberg, Germany}

\author{P.E. Filianin}
\affiliation{Max-Planck-Institut f\"ur Kernphysik, Saupfercheckweg 1, 69117 Heidelberg, Germany}
\affiliation{Physics Faculty of St.Petersburg State University, 198904, Peterhof, Russia}

\author{L. Gastaldo}
\affiliation{Kirchhoff Institut f\"ur Physik, Heidelberg Universit\"at, INF 227, 69120 Heidelberg, Germany}

\author{M. Goncharov}
\affiliation{Max-Planck-Institut f\"ur Kernphysik, Saupfercheckweg 1, 69117 Heidelberg, Germany}

\author{U. K\"oster}
\affiliation{Institut Laue-Langevin, 38042 Grenoble, France}

\author{F. Lautenschl\"ager}
\affiliation{GSI Helmholtzzentrum f\"ur Schwerionenforschung GmbH, Planckstra{\ss}e 1, 64291 Darmstadt, Germany}

\author{Yu.N. Novikov}
\affiliation{Max-Planck-Institut f\"ur Kernphysik, Saupfercheckweg 1, 69117 Heidelberg, Germany}
\affiliation{Physics Faculty of St.Petersburg State University, 198904, Peterhof, Russia}
\affiliation{Petersburg Nuclear Physics Institute, Gatchina, 188300 St. Petersburg, Russia}

\author{A. Rischka}
\affiliation{Max-Planck-Institut f\"ur Kernphysik, Saupfercheckweg 1, 69117 Heidelberg, Germany}

\author{R.X. Sch\"ussler}
\affiliation{Max-Planck-Institut f\"ur Kernphysik, Saupfercheckweg 1, 69117 Heidelberg, Germany}

\author{L. Schweikhard}
\affiliation{Institut f\"ur Physik, Ernst-Moritz-Arndt-Universit\"at, 17487 Greifswald, Germany}

\author{A. T\"urler}
\affiliation{Paul Scherrer Institute, 5232 Villigen, Switzerland}
\affiliation{Universit\"at Bern, 3012 Bern, Switzerland}

\begin{abstract}
The atomic mass difference of $^{163}$Ho and $^{163}$Dy has been directly measured with the Penning-trap  mass spectrometer SHIPTRAP applying the novel phase-imaging ion-cyclotron-resonance technique. Our measurement has solved the long-standing problem of large discrepancies in the $Q$-value of the electron capture in $^{163}$Ho determined by different techniques. Our measured mass difference shifts the current $Q$-value of 2555(16) eV evaluated in the AME2012 \citep{AME2012} by more than 7 sigma to 2833(30$_{stat}$)(15$_{sys}$) eV$/c^2$. With the new mass difference it will be possible, e.g., to reach in the first phase of the ECHo experiment a statistical sensitivity to the neutrino mass below 10 eV, which will reduce its present upper limit by more than an order of magnitude.  
\end{abstract}

% insert suggested PACS numbers in braces on next line
% insert suggested keywords - APS authors don't need to do this
%\keywords{}

%\maketitle must follow title, authors, abstract, \pacs, and \keywords
\maketitle

% body of paper here - Use proper section commands
% References should be done using the \cite, \ref, and \label commands

%\section{1}

One of the most interesting open questions in particle physics is the absolute scale of neutrino masses.
Among several approaches to determine the absolute neutrino masses, the analysis of the $\beta^-$ decays of tritium and $^{187}$Re and the electron capture (EC) in $^{163}$Ho are considered model-independent, since they are based on a kinematic analysis of the decay. The presently best upper limits of about 2.12 eV and 2.3 eV (95$\%$ C.L.) on the electron $antineutrino$ mass have been obtained in the "Troitsk $\nu$-mass" and "Neutrino Mainz" experiments (see \citep{Aseev,Kraus}), respectively, using the tritium $\beta^-$ decay. The best limit on the electron $neutrino$ mass, obtained by the analysis of the X-ray emission following the electron capture in $^{163}$Ho, is by far less stringent being about 225 eV (95$\%$ C.L.) \citep{Springer}.\\ 
Currently, several next-generation projects - KATRIN \citep{Drexlin} and Project 8 \citep{Project8} using tritium, MARE \citep{MARE} using $^{187}$Re, and ECHo \citep{Ranitzsch,ECHo}, HOLMES \citep{HOLMES} and NuMECS \citep{Numecs1,Numecs2} using $^{163}$Ho - are being developed with the goal to probe the electron-neutrino and antineutrino masses on a sub-eV level. In the kinematic analysis of the $\beta^-$ and EC spectra an accurate knowledge of the mass differences of the mother and daughter nuclides of the processes under investigation is essential for investigating systematic effects in the analysis of the endpoint region.\\
Presently, only high-precision Penning-trap mass spectrometry is capable of determining mass differences of nuclides relevant to the neutrino-mass determination with the required uncertainty (see, e.g., \citep{FSU,FSU2,Nesterenko}).\\  

\begin{figure*} [t]
\includegraphics[width=0.8\textwidth]{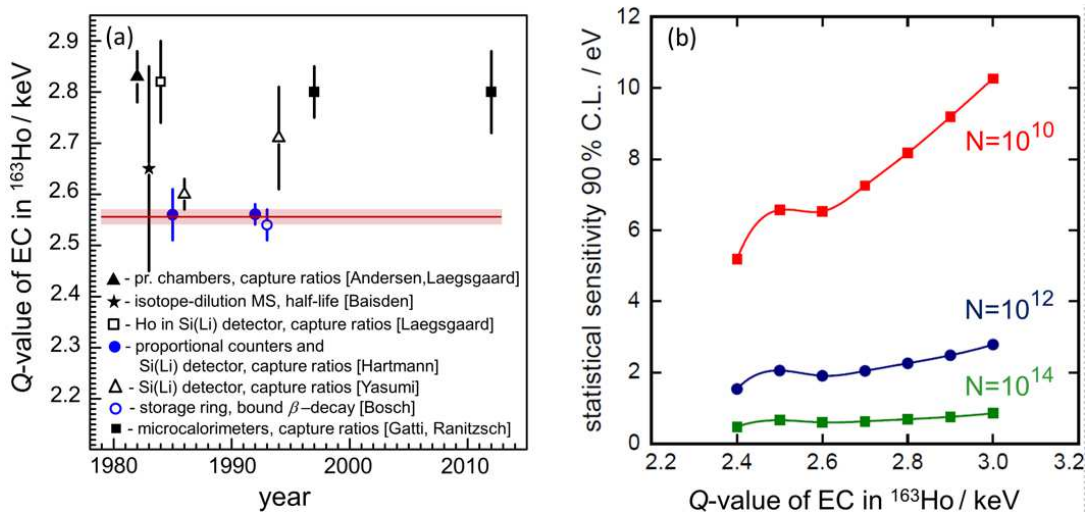}
\caption{\label{fig1} (color online) (a) The $Q$-value of the electron capture in $^{163}$Ho taken from \citep{AME2012} and obtained in several experiments from the analysis of the electron-capture spectrum (Andersen \citep{Andersen,Laegsgaard}, Baisden \citep{Baisden}, Laegsgaard \citep{Laegsgaard}, Hartmann \citep{Hartmann1,Hartmann2}, Yasumi \citep{Yasumi1,Yasumi2}, Bosch \citep{Bosch}, Gatti \citep{Gatti}, Ranitzsch \citep{Ranitzsch}) plotted according to the publication year. Different symbols indicate different experimental methods. The $Q$-value recently measured with TRIGA-TRAP \citep{TRIGA} is not shown in the plot due to its rather moderate accuracy of 700 eV \citep{TRIGA_Ho}. The red line and shaded band correspond to the recommended $Q$-value and its uncertainty, respectively \citep{AME2012}. The recommended $Q$-value was obtained by averaging only the data points which are colored blue in the plot. (b) Statistical sensitivity of the ECHo experiment \citep{ECHo} to the electron-neutrino mass as a function of the $Q$-value of the electron capture in $^{163}$Ho for several numbers $N$ of the acquired electron-capture events in the full energy spectrum (see text for details).}
\end{figure*}

In this Letter we report on the first direct high-precision Penning-trap determination of the atomic mass difference of $^{163}$Ho and $^{163}$Dy. The $Q$-value has already been determined, but only indirectly from the analysis of the EC-spectrum in several independent experiments by different groups using different methods (Fig.~\ref{fig1}(a)) \citep{Andersen,Baisden,Laegsgaard,Hartmann1,Yasumi1,Springer,Hartmann2,Bosch,Yasumi2,Gatti,Ranitzsch}.
The results fall in the range from approximately 2.4 keV to 2.9 keV, thus, exhibiting a substantial scatter of a few hundred eV. In particular, the $Q$-values obtained with cryogenic microcalorimetry \citep{Gatti,Ranitzsch} - the technique which forms the basis of all modern $^{163}$Ho-experiments - are higher by about 250 eV than the recommended $Q$-value of 2555(16) eV of the Atomic-Mass Evaluation AME2012 \citep{AME2012}, which was obtained by averaging only proportional counter data \citep{Hartmann1,Hartmann2} and storage-ring measurements \citep{Bosch}. Even if all the available values had been used for the averaging, the result would only slightly have been affected and still quite incompatible with the values obtained with cryogenic microcalorimetry. Recently, it has also been measured directly with the Penning-trap setup TRIGA-TRAP \citep{TRIGA}, however, with an uncertainty of 700 eV \citep{TRIGA_Ho}, which is insufficient to resolve the $Q$-value puzzle. If the recommended $Q$-value is correct, then the large deviation of the microcalorimetry values may be a sign of an insufficient understanding of the corresponding measurements of the EC spectrum, i.e. of the de-excitation processes involved in the EC in $^{163}$Ho. However, recent improved calculations of the probabilities of different atomic configurations in $^{163}$Dy after the EC in $^{163}$Ho \citep{Rob,Faes1,Faes2} including  2-hole and 3-hole excitations show that the contribution of higher order structures in the calorimetrically measured spectrum is below a few percent. Therefore, these higher orders cannot explain the large discrepancy between the result obtained by calorimetric measurements and the recommended value \citep{AME2012}.\\
\begin{figure*} [t]
\includegraphics[width=1\textwidth]{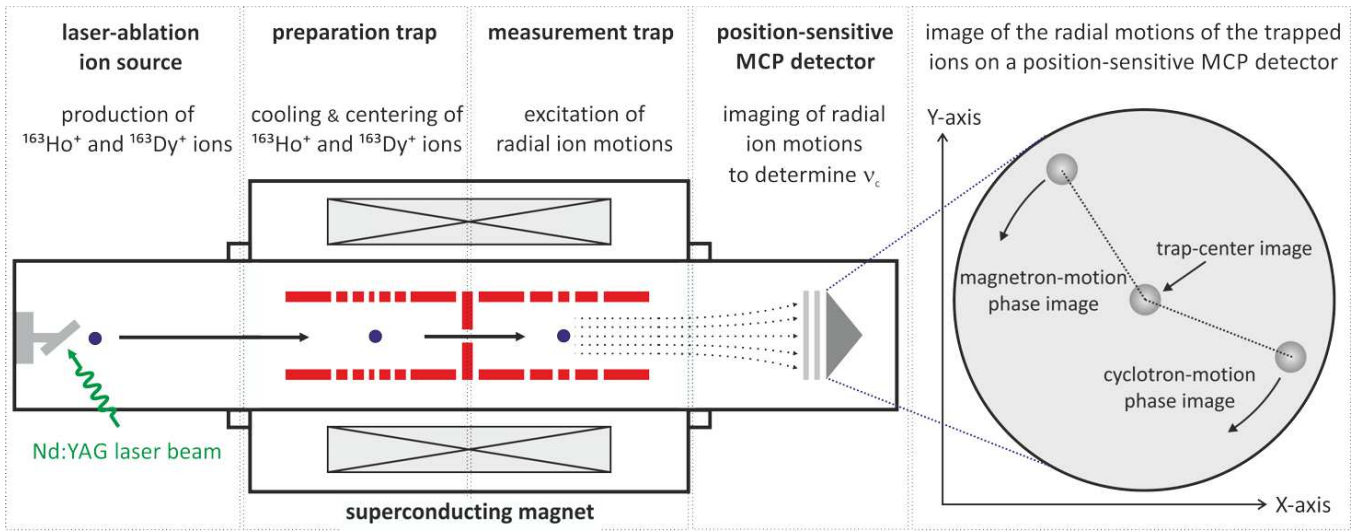}
\caption{\label{fig2} (color online) Schematic of the SHIPTRAP setup used for the determination of the $Q$-value of the electron capture in $^{163}$Ho. Note that while the ions perform cyclotron and magnetron revolutions in the same sense, their cyclotron phase image is inverted during the cyclotron-to-magnetron conversion \citep{PI-ICR-2}. For details see text, dimensions not to scale.}
\end{figure*} 
Furthermore, the statistical sensitivity of the experiments to the electron-neutrino mass value is a function of the $Q$-value of the EC in $^{163}$Ho.
Fig.~\ref{fig1}(b) shows the achievable statistical sensitivity (90$\%$ C.L.) of the ECHo experiment \citep{ECHo} to the electron-neutrino mass vs the $Q$-value for several numbers of acquired electron-capture events: a large uncertainty in the $Q$-value results in an unacceptably large uncertainty in the scale of the microcalorimetric experiment. Therefore, an accurate and independent direct measurement of the atomic mass differences of $^{163}$Ho and $^{163}$Dy is demanded.\\ 
   
The determination of the atomic mass difference of $^{163}$Ho and $^{163}$Dy was performed with the Penning-trap mass spectrometer SHIPTRAP \citep{Block-2007} by measuring the cyclotron-frequency ratio of $^{163}$Ho and $^{163}$Dy ions, $R=\nu_c(^{163}$Dy$^+)/\nu_c(^{163}$Ho$^+)$, using the novel Phase-Imaging Ion-Cyclotron Resonance (PI-ICR) technique \citep{PI-ICR-1,PI-ICR-2}. The cyclotron frequency $\nu_c$ of an ion with mass $m$ and charge $q$ in a magnetic field with strength $B$, given by $\nu_c=qB/(2\pi m)$, was determined as the sum of the two radial-motion frequencies of the trapped ions: magnetron frequency $\nu_-$ and modified cyclotron frequency $\nu_+$, i.e., $\nu_c$=$\nu_-$+$\nu_+$.\\       
A schematic of the experimental setup is presented in Fig.~\ref{fig2}.
Singly-charged ions of $^{163}$Ho and $^{163}$Dy were produced with a laser-ablation ion source \citep{Laser} by irradiating the corresponding Ho and Dy samples with a frequency-doubled Nd:YAG laser beam with a diameter of about 1 mm. This production mechanism of Ho-ions has already been demonstrated at the TRIGA-TRAP facility \citep{TRIGA_Ho}. For a production of the Dy-sample, a few milligrams of natural Dy in powder form were spread over a 5x5 mm$^2$ large titanium plate. $^{163}$Ho is radioactive with a half-life of 4570(25) years and thus first had to be produced in sufficient amount and in a high-purity form. The production of $^{163}$Ho involved neutron irradiation of an enriched $^{162}$Er sample in the high-flux reactor of the Institut Laue-Langevin and the subsequent electron capture decay of the resulting $^{163}$Er ($T_{1/2}=75$ min) into $^{163}$Ho. This was followed by a chemical separation based on ion-chromatography optimized to separate neighboring lanthanides. The resulting $^{163}$Ho contained less than 0.4$\%$ $^{163}$Dy - the only nuclide that cannot be resolved from $^{163}$Ho in the Penning trap and hence can lead to a systematic uncertainty in the mass difference determination between $^{163}$Ho and $^{163}$Dy. Finally, the Ho-sample for the laser ion source was prepared by putting a drop of $^{163}$Ho nitrate on a titanium plate and letting it dry. The final Ho-sample contained about 10$^{16}$ $^{163}$Ho atoms. The use of a sample with just a few micrograms of radioactive material for measuring the mass difference of heavy nuclides with a sub-ppb uncertainty is a unique feature of our experiment.\\
\begin{figure*} [t]
\includegraphics[width=1\textwidth]{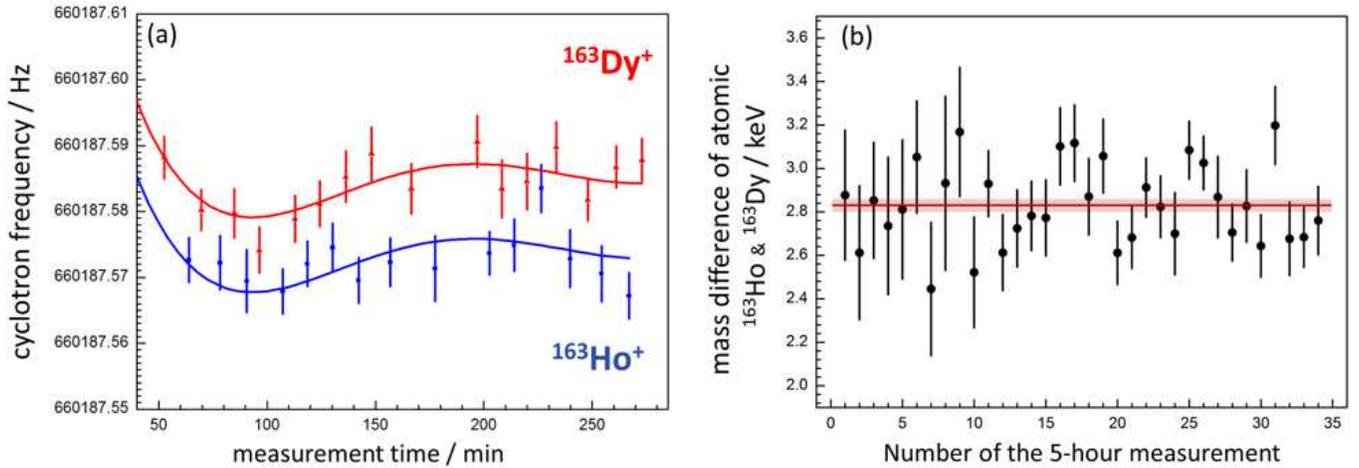}
\caption{\label{fig3} (color online) (a) An examplary 5-hour measurement period of the cyclotron frequencies $\nu_c$ of the $^{163}$Dy$^+$ and $^{163}$Ho$^+$ ions.  The ratio $R_{5 hour}$ of the cyclotron frequencies $\nu_c$ of the $^{163}$Dy$^+$ and $^{163}$Ho$^+$ ions was obtained along with the inner and outer errors \citep{Birge} by fitting to the $^{163}$Ho$^+$ frequency points a fifth order polynomial $P_1(t)$ and to the $^{163}$Dy$^+$ frequency points a polynomial $P_2(t)=R_{5 hour} \times P_1(t)$. (b) The mass difference of $^{163}$Ho and $^{163}$Dy calculated from the cyclotron-frequency ratios $R_{5 hour}$. The red line and the red shaded band are the average mass difference value and its uncertainty of the work reported here.}
\end{figure*}\\
From the laser-ablation ion source $^{163}$Ho$^+$ and $^{163}$Dy$^+$ ions were alternately transferred into a preparation trap for cooling and centering via mass-selective buffer-gas cooling \citep{Cooling} and further transfered into a measurement trap for cyclotron-frequency determination with the PI-ICR technique \citep{PI-ICR-1,PI-ICR-2}. The distance between the Ho and Dy samples on the target holder of the laser ion source was about 30 mm and thus a simultaneous irradiation of two samples and hence a simultaneous production of $^{163}$Ho and $^{163}$Dy ions were excluded. Other impurity ions were removed in the preparation trap with the buffer-gas cooling technique \citep{Cooling} prior to the transfer into the measurement trap. For the measurement of the ion cyclotron frequency "measurement scheme 2" as described in detail in \citep{PI-ICR-2} was applied:
in short, the amplitudes of the coherent components of their magnetron and axial motions were reduced to values of about 0.01 mm and 0.4 mm, respectively, by  simultaneously applying to the corresponding trap electrodes two 1-ms dipolar rf-pulses with certain amplitudes, initial phases and the corresponding frequencies. These steps were required to reduce to a level well below 10$^{-10}$ a possible shift in the cyclotron-frequency ratio of the $^{163}$Ho$^+$ and $^{163}$Dy$^+$ ions due to the anharmonicity of the trap potential and the inhomogeneity of the magnetic field. After these preparatory steps, the radius of the ion cyclotron motion was increased to 0.5 mm in order to set its initial phase of the cyclotron motion. Then, two excitation patterns, called in this work "magnetron phase" and "cyclotron phase", were applied alternately in order to measure the ion cyclotron frequency $\nu_c$. In the magnetron-phase pattern the cyclotron motion was first converted to the magnetron motion with the same radius. Then, the ions performed the magnetron motion accumulating a certain magnetron phase. After 600 ms elapsed, the ions' position in the trap plane perpendicular to the magnetic field was projected onto a position-sensitive detector by ejecting the ions from the trap towards the detector \citep{Eitel}. In the cyclotron-phase pattern the ions first performed the cyclotron motion for 600 ms accumulating the corresponding cyclotron phase with a consecutive  conversion to the magnetron motion and again projection of the ion position in the trap onto the position-sensitive detector. The angular FWHM of the magnetron and cyclotron phase spots with respect to the trap-image center amounts to about $7^0$ and $11^0$, respectively. The difference between the angular positions of the two phase images (see Fig.~\ref{fig2}) is a measure for the ion cyclotron frequency $\nu_c$.        
One measurement of the ion cyclotron frequency consisted of a periodic sequence of the magnetron and cyclotron pulse patterns with a period of about 800 ms and a total measurement time of approximately 5 minutes. On this time-scale and with the obtained uncertainty the phase measurements can be considered to be performed simultaneously. \\
Data with more than 5 detected ions (about ten loaded ions) per cycle were rejected in the data analysis in order
to reduce a possible shift in the cyclotron-frequency ratio of the $^{163}$Ho$^+$ and $^{163}$Dy$^+$ ions due to ion-ion interactions. To eliminate a cyclotron-frequency shift due to incomplete damping of the coherent component of the magnetron motion, the delay between the damping of the magnetron and axial motions and the excitation of the ion cyclotron motion was varied over the period of the magnetron motion. 
The positions of the magnetron and cyclotron phase spots were chosen such that the angle between them with respect to the measurement-trap axis did not exceed a few degrees. This procedure reduced the shift in the ratio of the $^{163}$Dy$^+$ and $^{163}$Ho$^+$ ions due to the possible distortion of the ion-motion projection onto the detector to a level well below 10$^{-10}$ \citep{PI-ICR-2}.\\
The cyclotron frequencies $\nu_c$ of the $^{163}$Dy$^+$ and $^{163}$Ho$^+$ ions were measured alternately for several days. The total measurement period was divided in 34 approximately 5-hour periods. For each of them the ratio $R_{5 hour}$ of the cyclotron frequencies $\nu_c$ of the $^{163}$Dy$^+$ and $^{163}$Ho$^+$ ions was obtained along with the inner and outer errors \citep{Birge} by simultaneously fitting a fifth-order polynomial to the $^{163}$Ho$^+$ frequency points and the same polynomial multiplied by a further fitted frequency ratio R$_{5 hour}$ to the $^{163}$Dy$^+$ frequency points (see Fig.~\ref{fig3}(a)).\\
The final cyclotron-frequency ratio $R$ is the weighted mean of the $R_{5 hour}$ ratios, where the inverse of the squared maxima of the inner and outer errors of the $R_{5 hour}$ ratios were taken as the weights to calculate $R$. The associated Birge ratio is 1.09.\\
\mbox{ } Fig.~\ref{fig3}(b) shows the mass difference of $^{163}$Ho and $^{163}$Dy corresponding to the cyclotron-frequency ratios $R_{5 hour}$.\\
The final frequency ratio $R$ with its statistical and systematic uncertainties as well as the corresponding mass difference of $^{163}$Ho and $^{163}$Dy are $R$=1.000$\,$000$\,$018$\,$67(20$_{stat}$)(10$_{sys}$) and $\Delta m$=2833(30$_{stat}$)(15$_{sys}$) eV/c$^2$, respectively. The systematic uncertainty in the frequency-ratio determination originates from the anharmonicity of the trap potential, the inhomogeneity of the magnetic field, the distortion of the ion-motion projection onto the detector, and a possible presence of $^{163}$Dy in the Ho-sample \citep{PI-ICR-2}.\\

Our result for the atomic mass difference of $^{163}$Ho and $^{163}$Dy deviates by more than seven sigma experimental uncertainty from the accepted value of the Atomic-Mass-Evaluation AME 2012 \citep{AME2012} while being in perfect agreement with the microcalorimetric measurements: $Q$ = 2800(50) eV \citep{Gatti} and $Q$ = 2800(80) eV \citep{Ranitzsch} (see Fig.~\ref{fig1}) - the $Q$-values, which were not included in the AME 2012 \citep{AME2012}. Thus, on the level of the present accuracy there are no unexpected deviations due to  systematic effects of cryogenic microcalorimetry or of the theoretical description of the spectrum. With the obtained $Q$-value and a foreseen number of acquired electron-capture events of $10^{10}$ in the first phase of the ECHo experiment (ECHo-1k) it will be possible to reach a statistical sensitivity below 10 eV to the neutrino mass, which will drastically, i.e., by more than an order of magnitude, improve the present upper limit on the neutrino mass.\\
For the determination of the electron-neutrino mass with sub-eV uncertainty, the $Q$-value must be determined with a substantially lower uncertainty, too. This independently measured $Q$-value on the eV level will remove any systematic uncertainties due to possible solid-state effects. Mass-difference measurements with correspondingly high accuracy will become possible with the realization of the PENTATRAP \citep{PENTATRAP-1,PENTATRAP-2} and CHIP-TRAP experiments \citep{CHIP}. Also the existing FSU-TRAP is in principle capable of determining the $Q$-value of the EC in $^{163}$Ho with an eV-uncertainty \citep{FSU}.\\

In summary, the atomic mass difference of $^{163}$Ho and $^{163}$Dy has been determined with the Penning-trap mass spectrometer SHIPTRAP with the novel PI-ICR technique. The measurement has yielded the value of 2833(30$_{stat}$)(15$_{sys}$) eV$/c^2$, in perfect agreement with the $Q$-values obtained with cryogenic microcalorimetry. It thus solves the puzzle in the determination of the $Q$-value in the EC in $^{163}$Ho and allows for defining the scale of the experiments on the determination of the electron-neutrino mass from the electron capture in $^{163}$Ho.\\

% If you have acknowledgments, this puts in the proper section head.
%\begin{acknowledgments}
This research was performed in the framework of the DFG Research Unit FOR 2202 "Neutrino Mass Determination by Electron Capture in Holmium-163 – ECHo" and was supported by the Max-Planck Society, by the EU (ERC grant number 290870-MEFUCO), by the German BMBF (05P12HGFN5, 05P12HGFNE and 01DJ14002) and by the Russian Ministry of Education and Science (project 2.2). 
%\end{acknowledgments}

% Create the reference section using BibTeX:
% see BibTeX file GdSm.bib
\bibliography{Ho}

\end{document}